\documentclass[10pt,a4paper]{article}
\usepackage{amsmath,amssymb,amsfonts}
\usepackage{graphicx}
\usepackage{hyperref}
\usepackage[margin=0.75in]{geometry}
\usepackage{float}
\usepackage{bm}
\usepackage{booktabs}
\usepackage{tabularx}
\usepackage{subcaption}

\usepackage{setspace}
\usepackage{titlesec}
\titlespacing*{\section}{0pt}{8pt}{4pt}
\titlespacing*{\subsection}{0pt}{6pt}{3pt}

\AtBeginDocument{
  \setlength{\abovedisplayskip}{4pt}
  \setlength{\belowdisplayskip}{4pt}
  \setlength{\abovedisplayshortskip}{2pt}
  \setlength{\belowdisplayshortskip}{2pt}
}

\let\oldbibliography\thebibliography
\renewcommand{\thebibliography}[1]{%
  \oldbibliography{#1}%
  \setlength{\itemsep}{0pt}%
  \setlength{\parskip}{0pt}%
}

\title{Three-Dimensional Variational Data Assimilation with Rapid Update Cycling for Short-Range Precipitation Forecasting: A Case Study of Heavy Rainfall in Bali, Indonesia}

\author{
Nurjanna J. Trilaksono$^{1,*}$,
Sandy H. S. Herho$^{2,3}$,
I Putu F. Wistika$^{1}$,\\
Faiz R. Fajary$^{1}$,
Rusmawan Suwarman$^{1}$,
and Dasapta E. Irawan$^{4}$
}

\date{}

\begin{document}
\maketitle

\begin{center}
\small
$^{1}$Atmospheric Science Research Group, Bandung Institute of Technology (ITB), Jalan Ganesha 10, Bandung, 40132, West Java, Indonesia\\
$^{2}$Ronin Institute for Independent Scholarship, 2108 N St STE N, Sacramento, CA 95816, USA\\
$^{3}$School of Systems Science and Industrial Engineering, State University of New York (SUNY), 4400 Vestal Parkway East, Binghamton, NY 13902, USA\\
$^{4}$Applied Geology Research Group, Bandung Institute of Technology (ITB), Jalan Ganesha 10, Bandung, 40132, West Java, Indonesia\\
$^{*}$Correspondence: jpatiani@itb.ac.id
\end{center}

\begin{abstract}
\noindent
This study evaluates the effectiveness of three-dimensional variational (3D-Var) data assimilation coupled with a Rapid Update Cycle (RUC) framework for improving short-range precipitation forecasts over the Indonesian Maritime Continent (IMC). We employ the Weather Research and Forecasting (WRF) model and its data assimilation component (WRFDA) to assimilate surface observations from Automatic Weather Stations (AWS) at cycling intervals of 1, 3, 6, and 12~hours. Our test case is a heavy rainfall event on 7~July 2023 in Bali Province, during which accumulated precipitation exceeded 193~mm~day$^{-1}$. The 1-hour cycling interval yields the lowest root-mean-square error (RMSE) for both 2-meter temperature (0.0--0.3$\,^\circ$C) and hourly precipitation (1.295~mm~h$^{-1}$), corresponding to reductions of roughly 75\% and 57\%, respectively, relative to non-assimilated forecasts. Frequent cycling constrains initial-condition errors and captures mesoscale convective evolution, as confirmed by improved spatial agreement with radar reflectivity observations. These results demonstrate that high-frequency assimilation cycling offers clear advantages for nowcasting in tropical maritime environments.
\end{abstract}

\noindent\textbf{Keywords:} data assimilation; numerical weather prediction; rapid update cycle; tropical meteorology; variational methods

%===================================================================
\section{Introduction}

The accuracy of numerical weather prediction (NWP) hinges on the quality of the initial conditions from which the atmospheric state is integrated forward in time~\cite{r01}. Because the governing equations of atmospheric motion---the Navier--Stokes momentum equations, the mass continuity equation, and the thermodynamic energy equation---define an initial value problem, even small errors in the starting state can amplify rapidly through nonlinear dynamics~\cite{r02}. Lorenz~\cite{r03} was the first to characterize this sensitive dependence, and his work underscores why accurate initialization remains one of the central challenges in weather forecasting.

Data assimilation (DA) tackles this challenge by blending a prior model forecast (the background, or first guess) with available observations to produce an improved atmospheric state estimate known as the analysis. Among DA techniques, the variational approach minimizes a cost function that penalizes discrepancies from both the background and observations, each weighted by their respective error covariances~\cite{r05,r06}. Three-dimensional variational (3D-Var) DA, in particular, has become a workhorse of operational NWP owing to its computational efficiency and well-understood theoretical basis~\cite{r07,r08}.

Forecasting precipitation over the Indonesian Maritime Continent (IMC) poses distinctive difficulties. The region's archipelagic topography, persistently warm sea surface temperatures, and strong diurnal convective forcing give rise to precipitation systems that are highly localized and evolve on timescales of just a few hours~\cite{r09}. Earlier work has shown that Weather Research and Forecasting (WRF) model simulations over Bali Province tend to underestimate observed rainfall, a bias linked in part to deficiencies in the initial conditions~\cite{r10}.

The Rapid Update Cycle (RUC) methodology---which performs DA at sub-synoptic intervals---offers a way to rein in forecast drift by updating the model state more frequently with fresh observations~\cite{r11,r12}. Operationally, the RUC concept has been in use at the National Centers for Environmental Prediction (NCEP) since the 1990s~\cite{r13}. More recently, Chen et~al.~\cite{r14} demonstrated that hourly 3D-Var cycling in China improved precipitation forecast skill by nearly 100\% relative to conventional 6- or 12-hourly cycling. Whether similar gains are achievable over the IMC, however, has not been tested. The present study fills this gap by systematically comparing cycling intervals of 1, 3, 6, and 12~hours for a heavy-rainfall case in Bali, and by quantifying the degree to which more frequent assimilation translates into better short-range forecasts in this tropical maritime setting.

\section{Mathematical Framework}

The goal of 3D-Var DA is to find an analysis state $\mathbf{x}^{a}$, a vector in $\mathbb{R}^{n}$ (where $n$ is the dimension of the model state), that best fits both a background state $\mathbf{x}^{b}$ and a vector of observations $\mathbf{y}^{o} \in \mathbb{R}^{p}$ (where $p$ is the number of observations). This is achieved by minimizing the quadratic cost function~\cite{r05,r06}
\begin{equation}
J(\mathbf{x}) = \frac{1}{2}(\mathbf{x} - \mathbf{x}^{b})^{\mathrm{T}} \mathbf{B}^{-1} (\mathbf{x} - \mathbf{x}^{b}) + \frac{1}{2}\bigl(\mathbf{y}^{o} - H(\mathbf{x})\bigr)^{\mathrm{T}} \mathbf{R}^{-1} \bigl(\mathbf{y}^{o} - H(\mathbf{x})\bigr),
\label{eq:cost}
\end{equation}
in which $\mathbf{B} \in \mathbb{R}^{n \times n}$ denotes the background error covariance matrix, $\mathbf{R} \in \mathbb{R}^{p \times p}$ is the observation error covariance matrix, and $H\colon \mathbb{R}^{n} \to \mathbb{R}^{p}$ is the observation operator that maps the model state into observation space. The first term on the right-hand side of Eq.~(\ref{eq:cost}) penalizes departures from the background, weighted by $\mathbf{B}^{-1}$; the second penalizes departures from the observations, weighted by $\mathbf{R}^{-1}$.

Setting the gradient of $J$ to zero at the analysis gives the optimality condition
\begin{equation}
\nabla J(\mathbf{x}^{a}) = \mathbf{B}^{-1}(\mathbf{x}^{a} - \mathbf{x}^{b}) - \mathbf{H}^{\mathrm{T}} \mathbf{R}^{-1}\bigl(\mathbf{y}^{o} - H(\mathbf{x}^{a})\bigr) = \mathbf{0},
\label{eq:gradient}
\end{equation}
where $\mathbf{H} = \partial H / \partial \mathbf{x}$ is the Jacobian of the observation operator (i.e., its linearization). When $H$ is linear, $H(\mathbf{x}) = \mathbf{H}\mathbf{x}$, the analysis has the well-known closed-form solution
\begin{equation}
\mathbf{x}^{a} = \mathbf{x}^{b} + \mathbf{K}\bigl(\mathbf{y}^{o} - \mathbf{H}\mathbf{x}^{b}\bigr),
\label{eq:analysis}
\end{equation}
where $\mathbf{K} = \mathbf{B}\mathbf{H}^{\mathrm{T}}(\mathbf{H}\mathbf{B}\mathbf{H}^{\mathrm{T}} + \mathbf{R})^{-1}$ is the gain matrix that distributes observational increments across the model grid. The quantity $\mathbf{y}^{o} - \mathbf{H}\mathbf{x}^{b}$, called the innovation, measures the mismatch between the observations and the background projected into observation space.

Working directly with $\mathbf{B}$ is impractical in operational NWP, where $n$ can reach $\sim\!10^{7}$~\cite{r07}. The WRF Data Assimilation (WRFDA) system circumvents this by introducing a control variable transform $\mathbf{x}' = \mathbf{U}\mathbf{v}$, with $\mathbf{x}' = \mathbf{x} - \mathbf{x}^{b}$ and $\mathbf{U}$ chosen so that $\mathbf{B} = \mathbf{U}\mathbf{U}^{\mathrm{T}}$~\cite{r08}. In terms of the new variable $\mathbf{v}$, the cost function becomes
\begin{equation}
J(\mathbf{v}) = \frac{1}{2}\mathbf{v}^{\mathrm{T}}\mathbf{v} + \frac{1}{2}\bigl(\mathbf{y}^{o} - H(\mathbf{x}^{b} + \mathbf{U}\mathbf{v})\bigr)^{\mathrm{T}} \mathbf{R}^{-1} \bigl(\mathbf{y}^{o} - H(\mathbf{x}^{b} + \mathbf{U}\mathbf{v})\bigr),
\label{eq:cost_cv}
\end{equation}
which is minimized iteratively using a conjugate-gradient algorithm~\cite{r15}.

In WRFDA, the transform $\mathbf{U}$ is factored as $\mathbf{U} = \mathbf{U}_{p}\,\mathbf{U}_{v}\,\mathbf{U}_{h}$. Here $\mathbf{U}_{h}$ encodes horizontal error correlations through recursive filters whose length scales are tunable, $\mathbf{U}_{v}$ captures vertical correlations via empirical orthogonal function (EOF) decomposition, and $\mathbf{U}_{p}$ enforces dynamical balance constraints among the control variables~\cite{r08,r16}. The underlying background error statistics are estimated with the NMC method of Parrish and Derber~\cite{r06}, which approximates forecast error covariances from differences between pairs of forecasts at different lead times that verify at the same valid time.

The RUC embeds 3D-Var within an intermittent cycling loop. Denote the sequence of analysis times by $t_{0}, t_{1}, \ldots, t_{N}$, separated by a cycling interval $\Delta t_{c}$. At each time $t_{k}$ two steps are carried out. First, the \textit{analysis step}: given a background field $\mathbf{x}^{b}_{t_{k}}$ (from a cold start at $k=0$, or from the previous cycle's short-range forecast otherwise) together with observations $\mathbf{y}^{o}_{t_{k}}$, solve Eq.~(\ref{eq:cost_cv}) to obtain the analysis $\mathbf{x}^{a}_{t_{k}}$. Second, the \textit{forecast step}: integrate the full nonlinear WRF model $\mathcal{M}$ forward in time:
\begin{equation}
\mathbf{x}^{b}_{t_{k+1}} = \mathcal{M}_{t_{k} \to t_{k+1}}\!\bigl(\mathbf{x}^{a}_{t_{k}}\bigr).
\label{eq:forecast}
\end{equation}
Because each cycle's background is spawned from the previous cycle's analysis, error corrections accumulate and the model's hydrometeorological fields have time to spin up---an important practical advantage over cold-start re-initialization from global analyses~\cite{r14,r18}.

Forecast quality is assessed with two standard scalar scores. The root-mean-square error (RMSE) is defined as
\begin{equation}
\mathrm{RMSE} = \sqrt{\frac{1}{N}\sum_{i=1}^{N}(F_{i} - O_{i})^{2}},
\label{eq:rmse}
\end{equation}
where $F_{i}$ and $O_{i}$ are the forecast and observed values at the $i$-th verification point and $N$ is the number of such points. The mean bias is
\begin{equation}
\mathrm{Bias} = \frac{1}{N}\sum_{i=1}^{N}(F_{i} - O_{i}).
\label{eq:bias}
\end{equation}
Together, RMSE and Bias characterize the random and systematic components of forecast error.

\section{Numerical Implementation}

\begin{figure}[H]
\centering
\includegraphics[width=0.7\textwidth]{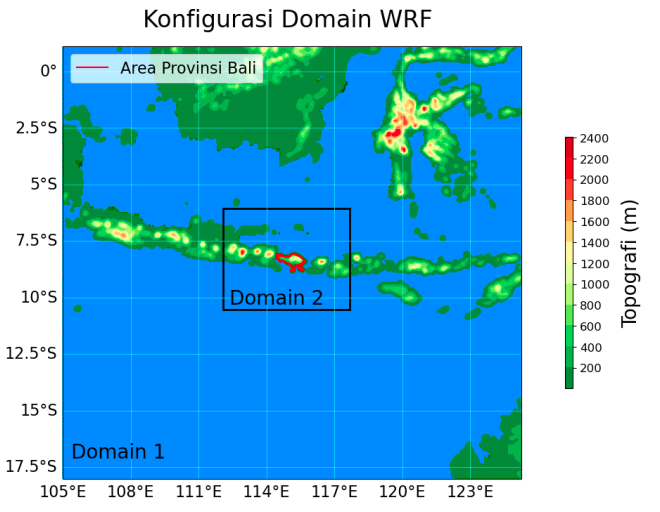}
\caption{Computational domain configuration. Domain~1 (outer) has 9~km horizontal resolution; Domain~2 (inner, rectangle) has 3~km resolution. Shading shows topographic elevation (m). Bali Province lies within Domain~2.}
\label{fig:domain}
\end{figure}

We use the WRF model version~4.4~\cite{r19} in a two-domain nested configuration centered on Bali Province, Indonesia (Fig.~\ref{fig:domain}). The outer domain (Domain~1) spans $113$--$117^\circ$E and $7^\circ$N--$9^\circ$S at 9~km horizontal grid spacing; the inner domain (Domain~2) covers $114.15$--$115.45^\circ$E and $8^\circ$N--$9^\circ$S at 3~km resolution. This arrangement allows the outer domain to capture synoptic-scale forcing while the inner domain resolves mesoscale convective organization. Vertical levels follow terrain-following eta ($\eta$) coordinates with a model top at 50~hPa.

Table~\ref{tab:physics} lists the physics parameterizations. Microphysics uses the WRF Single-Moment 3-class (WSM3) scheme, which carries prognostic variables for water vapor, cloud water/ice, and rain/snow~\cite{r25}. Longwave radiation follows the Rapid Radiative Transfer Model (RRTM)~\cite{r26}, and shortwave radiation uses the Dudhia scheme~\cite{r27}. The surface layer is based on Monin--Obukhov similarity theory~\cite{r28}, the planetary boundary layer (PBL) on the Medium-Range Forecast (MRF) scheme~\cite{r29}, and the land surface on the thermal diffusion approach. Cumulus convection is parameterized on Domain~1 with the Grell--D\'{e}v\'{e}nyi ensemble scheme~\cite{r30}; Domain~2, at 3~km, treats convection explicitly.

\begin{table}[H]
\caption{WRF model physics parameterization schemes}\small\smallskip
\centering
\begin{tabular}{lll}
\toprule
\textbf{Physical Process} & \textbf{Scheme} & \textbf{Reference} \\
\midrule
Microphysics & WSM 3-class simple ice & Hong et al.~\cite{r25} \\
Longwave radiation & RRTM & Mlawer et al.~\cite{r26} \\
Shortwave radiation & Dudhia & Dudhia~\cite{r27} \\
Surface layer & Monin--Obukhov similarity & Paulson~\cite{r28} \\
Land surface & Thermal diffusion & Skamarock et al.~\cite{r19} \\
Planetary boundary layer & MRF & Hong and Pan~\cite{r29} \\
Cumulus (Domain 1 only) & Grell--D\'{e}v\'{e}nyi ensemble & Grell and D\'{e}v\'{e}nyi~\cite{r30} \\
\bottomrule
\end{tabular}
\label{tab:physics}
\end{table}

Initial and lateral boundary conditions come from the Global Forecast System (GFS) operational analysis at $0.25^\circ$ horizontal and 3-hourly temporal resolution. Within WRFDA we use the 3D-Var method with control variable option CV3, which relies on domain-independent global background error statistics distributed with WRFDA. We also generate domain-specific background error covariances with the \texttt{gen\_be} utility, following the NMC method~\cite{r06}: differences between 24-hour and 12-hour forecasts valid at the same time are accumulated over 1--31~July 2023, using forecast pairs initialized at 00~UTC and 12~UTC daily.

\begin{figure}[H]
\centering
\includegraphics[width=0.6\textwidth]{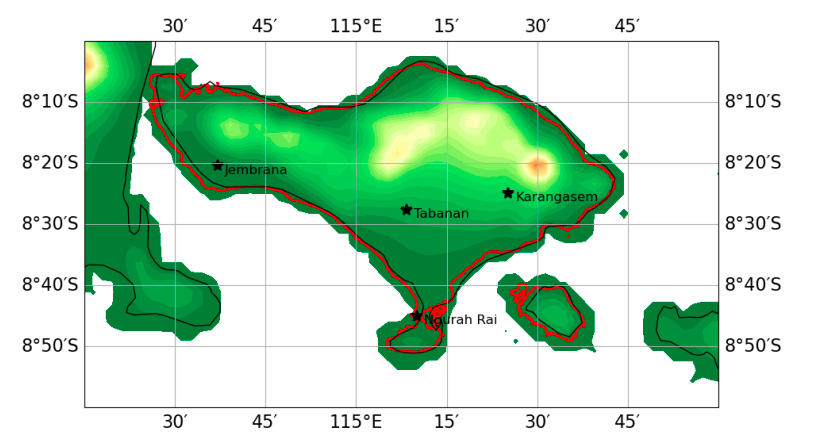}
\caption{Locations of the five AWS sites used for DA and verification. Topographic elevation (m) is shown by shading.}
\label{fig:aws}
\end{figure}

Surface observations from five AWS sites distributed across Bali Province supply the observational constraint (Fig.~\ref{fig:aws}). The stations---in Jembrana, Tabanan, Gianyar, Karangasem, and at Ngurah Rai International Airport---record temperature ($T$), relative humidity (RH), surface pressure ($p_{s}$), and wind speed/direction at 10-minute intervals. Raw data are preprocessed with the OBSPROC utility, which handles quality control, temporal interpolation to analysis times, and format conversion to LITTLE\_R for WRFDA ingestion. Prescribed observation errors follow standard practice: 1~K for $T$, 10\% for RH, 1~hPa for $p_{s}$, and 1~m~s$^{-1}$ for each wind component.

\begin{figure}[H]
\centering
\includegraphics[width=0.85\textwidth]{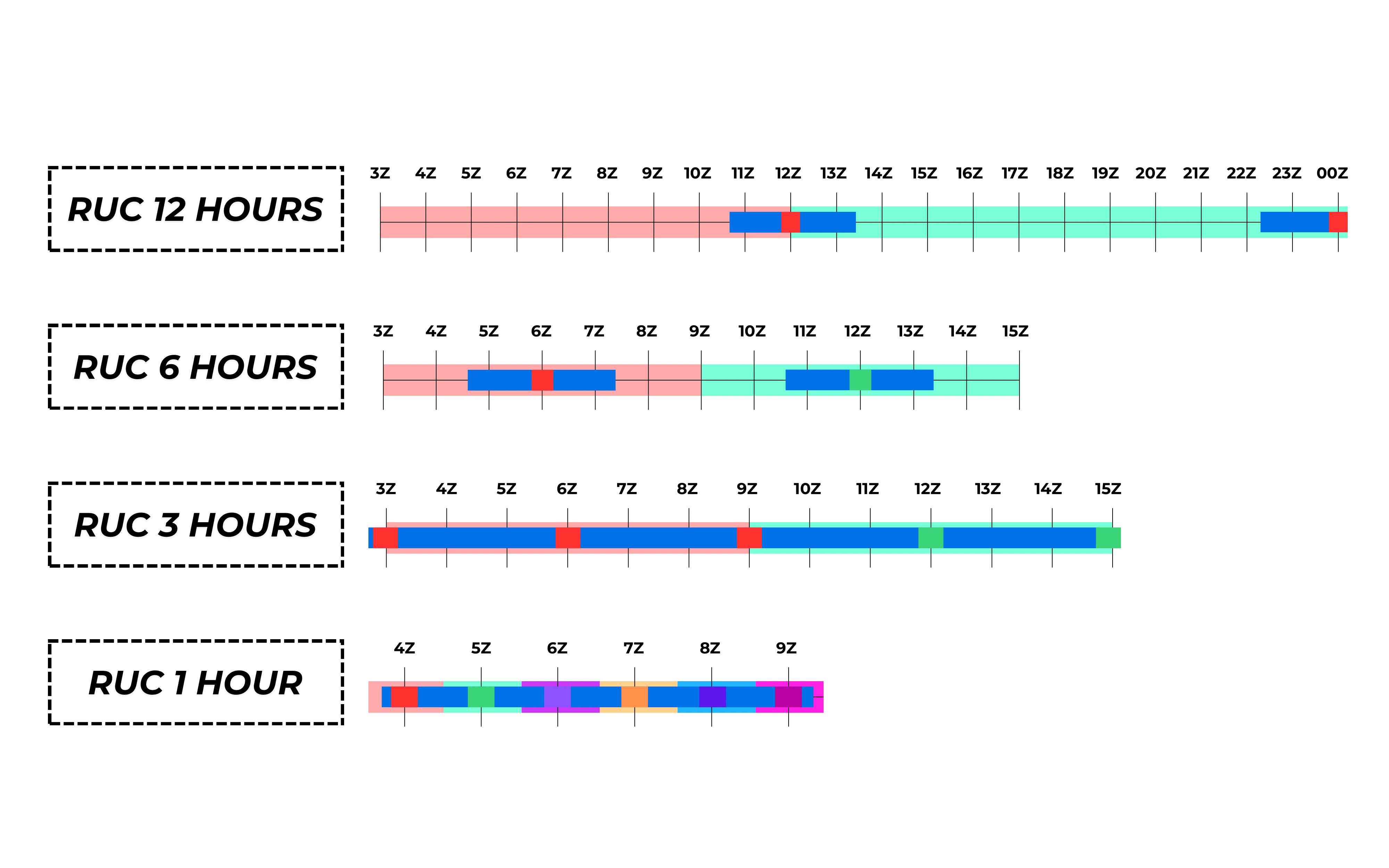}
\caption{Schematic of the RUC experimental design. Colored blocks mark analysis times; horizontal lines denote forecast integrations. The four cycling configurations use intervals of 12, 6, 3, and 1~hours. The verification target time is 08:00~UTC (16:00~WITA) on 7~July 2023.}
\label{fig:ruc}
\end{figure}

Six experiments test the sensitivity to cycling frequency (Fig.~\ref{fig:ruc}). The control run (Non-DA) uses GFS initial conditions directly, with no DA. The DA experiment performs a single 3D-Var analysis at the initial time only. Four RUC experiments---Cycle~12, Cycle~6, Cycle~3, and Cycle~1---assimilate observations at 12-, 6-, 3-, and 1-hourly intervals, respectively, using the warm-start cycling procedure. All runs start at 08:00 Central Indonesian Time (WITA; 00:00~UTC) on 6~July 2023 and run through 08:00~WITA on 8~July 2023, giving 48-hour forecasts.

On 7~July 2023, widespread convective activity triggered flooding across western and central Bali. The rain gauge at Pos Meliling Kerambitan in Tabanan Regency recorded 193.5~mm~day$^{-1}$---well above the 50~mm~day$^{-1}$ extreme-rainfall threshold defined by the Indonesian Agency for Meteorology, Climatology, and Geophysics (BMKG). The heaviest hourly rates occurred near 16:00~WITA (08:00~UTC), coinciding with the afternoon peak of tropical diurnal convection. We verify model output against AWS observations and composite reflectivity from the BMKG Ngurah Rai weather radar at this time.

\section{Results}

\begin{figure}[H]
\centering
\includegraphics[width=0.7\textwidth]{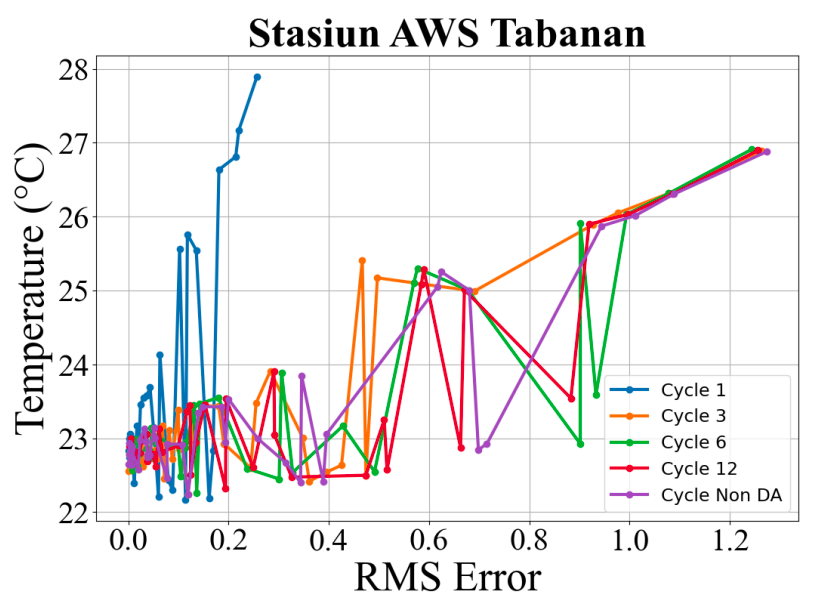}
\caption{RMSE of 2-meter temperature ($^\circ$C) at the Tabanan AWS for all experiments, 6~July 08:00 to 8~July 08:00~WITA. Lower values indicate closer agreement with observations.}
\label{fig:rmse_temp}
\end{figure}

Figure~\ref{fig:rmse_temp} shows the RMSE of 2-meter temperature at the Tabanan AWS, computed from hourly model output over the full 48-hour window (6~July 08:00~WITA to 8~July 08:00~WITA). Cycle~1 stands out clearly, with RMSE confined to 0.0--0.3$\,^\circ$C throughout the period. Accuracy degrades monotonically as cycling becomes less frequent: Cycle~3 produces RMSE of 0.2--0.4$\,^\circ$C, Cycle~6 reaches $\sim$0.8$\,^\circ$C, and Cycle~12 approaches 1.1$\,^\circ$C. The Non-DA run fares worst, with errors exceeding 1.2$\,^\circ$C---evidence that the unassimilated forecast drifts substantially from observed conditions over two days.

Table~\ref{tab:rmse} summarizes the RMSE of hourly precipitation at two verification sites: Tabanan (western Bali) and Denpasar (central Bali). At Tabanan, Cycle~1 achieves an RMSE of 1.295~mm~h$^{-1}$, a 57\% reduction compared with Non-DA (3.023~mm~h$^{-1}$). The improvement is monotonic: Cycle~3 gives 1.432~mm~h$^{-1}$, Cycle~6 gives 1.745~mm~h$^{-1}$, and Cycle~12 gives 2.572~mm~h$^{-1}$. A similar pattern holds at Denpasar, where Cycle~1 (1.487~mm~h$^{-1}$) cuts the Non-DA error (3.954~mm~h$^{-1}$) by 62\%.

\begin{table}[H]
\caption{RMSE of hourly precipitation (mm~h$^{-1}$) at AWS verification sites, 6~July 08:00 to 8~July 08:00~WITA}\small\smallskip
\centering
\begin{tabular}{lcccccc}
\toprule
\textbf{Station} & \textbf{Cycle~1} & \textbf{Cycle~3} & \textbf{Cycle~6} & \textbf{Cycle~12} & \textbf{DA} & \textbf{Non-DA} \\
\midrule
AWS Tabanan       & 1.295 & 1.432 & 1.745 & 2.572 & 2.482 & 3.023 \\
Geofisika Denpasar & 1.487 & 1.853 & 1.855 & 3.224 & 3.877 & 3.954 \\
\bottomrule
\end{tabular}
\label{tab:rmse}
\end{table}

\begin{figure}[H]
\centering
\includegraphics[width=0.7\textwidth]{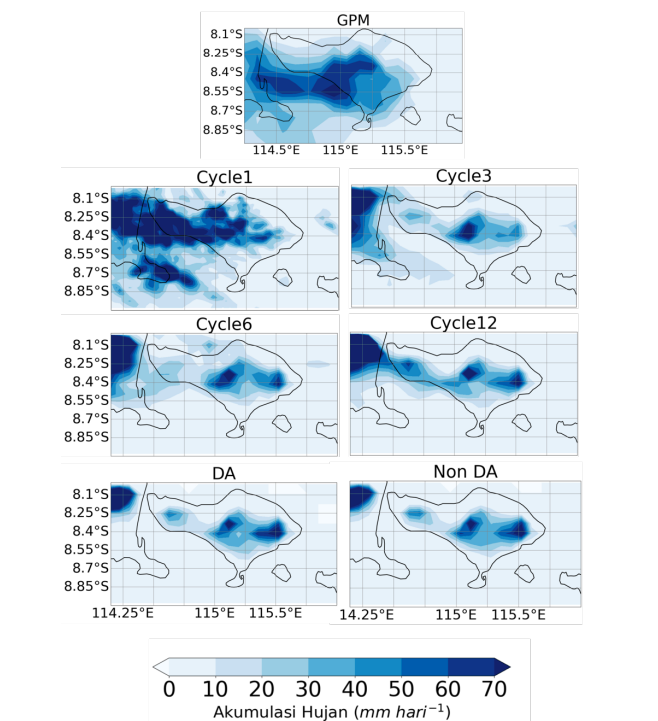}
\caption{Accumulated precipitation (mm~day$^{-1}$) for 7~July 2023 (00:00--23:00~WITA). Top: GPM satellite estimate. Remaining panels: Cycle~1, Cycle~3, Cycle~6, Cycle~12, DA, and Non-DA.}
\label{fig:precip}
\end{figure}

Figure~\ref{fig:precip} compares accumulated rainfall on 7~July 2023 (00:00--23:00~WITA) across all experiments with satellite estimates from the Global Precipitation Measurement (GPM) mission. GPM shows rainfall concentrated over western and southwestern Bali, with secondary maxima over the central highlands. Cycle~1 reproduces this pattern most faithfully---both the location of the precipitation maxima and the east-west gradient are captured. With progressively coarser cycling (Cycle~3, 6, 12), the modeled rainfall shifts eastward toward central Bali and the western totals are increasingly underestimated. The Non-DA experiment displaces precipitation most severely, placing the bulk of rainfall over central Bali while largely missing the heavy accumulations observed in the west.

\begin{figure}[H]
\centering
\includegraphics[width=0.7\textwidth]{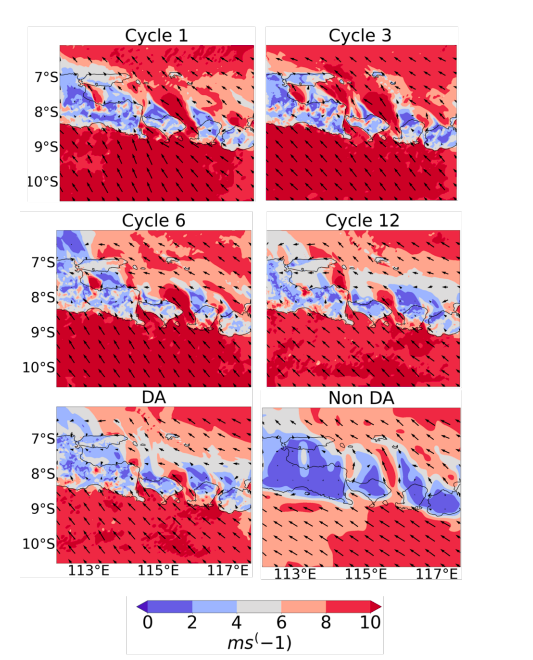}
\caption{Bias of 10-meter wind speed (m~s$^{-1}$) at the Tabanan AWS, 6~July 08:00 to 7~July 16:00~WITA. Negative values denote model underestimation.}
\label{fig:wind_bias}
\end{figure}

Wind bias provides a dynamical perspective on the precipitation differences. Figure~\ref{fig:wind_bias} presents the bias of 10-meter wind speed at the Tabanan AWS. All experiments underestimate wind speed, but the magnitude shrinks with more frequent cycling. Cycle~1 shows a bias near $-1.25$~m~s$^{-1}$; this grows to $-1.75$~m~s$^{-1}$ for Cycle~3, $-2.5$~m~s$^{-1}$ for Cycle~6, $-3.0$~m~s$^{-1}$ for Cycle~12, and $-4.5$~m~s$^{-1}$ for Non-DA. The smaller wind bias in the higher-frequency experiments points to a better representation of low-level convergence---the very mechanism that organizes and sustains tropical convection.

\begin{figure}[H]
\centering
\includegraphics[width=0.7\textwidth]{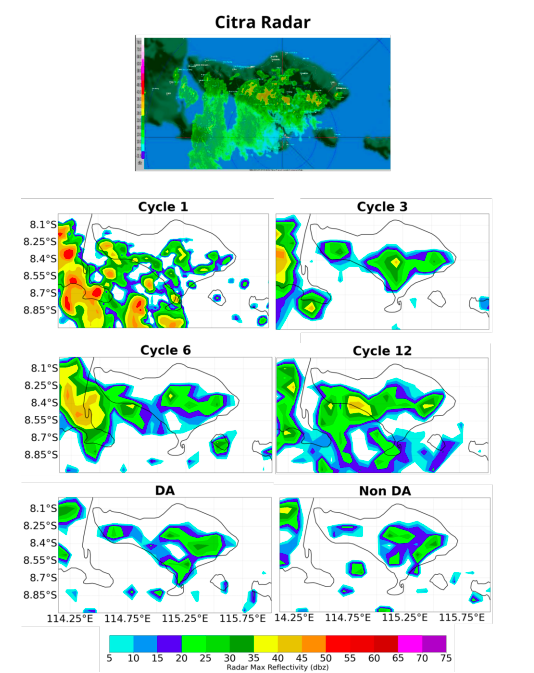}
\caption{Maximum radar reflectivity (dBZ) at 16:00~WITA, 7~July 2023. Top: BMKG Ngurah Rai radar observation. Remaining panels: simulated reflectivity from Cycle~1, Cycle~3, Cycle~6, Cycle~12, DA, and Non-DA.}
\label{fig:radar}
\end{figure}

An independent check against weather radar observations further supports the cycling-frequency findings. Figure~\ref{fig:radar} juxtaposes the BMKG Ngurah Rai composite reflectivity with model-derived maximum reflectivity at 16:00~WITA on 7~July 2023. The radar image reveals widespread convection with embedded cores above 40~dBZ across western and central Bali. Cycle~1 yields reflectivity of 20--40~dBZ distributed over the same convective areas---a reasonable match. Cycle~3 and Cycle~6 produce weaker echoes (10--25~dBZ), confined too narrowly to central Bali. Non-DA generates only 5--10~dBZ, entirely failing to capture the observed convective intensity.

\section{Discussion}

The monotonic improvement in forecast skill with increasing cycling frequency seen here is consistent with theoretical expectations and with operational experience elsewhere~\cite{r11,r13,r14}. Each assimilation cycle corrects initial-condition errors before they can amplify through nonlinear instabilities, so more cycles mean less drift. At a 1-hour interval, the cycling period approaches the intrinsic decorrelation timescale of mesoscale convective systems in the tropics~\cite{r20}, allowing the analysis to ``keep up'' with rapidly evolving weather that coarser cycling would miss. The 57\% reduction in precipitation RMSE between Cycle~1 and Non-DA is at the high end of what has been reported in midlatitude studies, though it is in line with the Chinese cycling experiments of Chen et~al.~\cite{r14}.

Two features of the tropical maritime environment help explain why frequent cycling pays off so handsomely here. First, the IMC is dominated by diurnal convective forcing that spawns organized precipitation systems on timescales of hours---well below the 6--12-hour cycling intervals that remain standard in many operational centers~\cite{r09}. Second, the conventional observation network in the Indonesian archipelago is sparse, so each AWS datum carries disproportionate weight; making full use of every available observation is therefore critical~\cite{r21}.

The warm-start strategy also plays an important role. Because each cycle inherits hydrometeorological fields from the previous cycle's forecast rather than starting from scratch with interpolated GFS analyses, the model can produce precipitation almost immediately after initialization~\cite{r18}. Cold starts, by contrast, need several hours of integration before microphysical fields spin up, degrading the earliest forecast hours. This spin-up penalty is most acute at convective scales, where precipitation develops quickly once the environment becomes favorable.

Several caveats apply. We analyze a single heavy-rainfall event; generalizing the conclusions requires testing across a wider range of cases and seasons. The 3D-Var method used here relies on static, climatological background error covariances, which do not adapt to flow-dependent error structures~\cite{r22}. Hybrid ensemble--variational schemes that blend ensemble-derived covariances into the 3D-Var framework could yield further improvements, particularly at convective scales~\cite{r23}. The five-station AWS network, while adequate for a proof of concept, provides limited spatial coverage; direct assimilation of radar reflectivity and radial velocity into WRFDA would substantially strengthen the mesoscale constraint~\cite{r24}. Finally, the computational overhead of hourly cycling deserves mention. Performing 24 assimilation cycles per day costs roughly 24 times as much DA computation as a single daily analysis. In practice, however, the WRF forecast integration dominates the total wall-clock time, and the shorter forecast horizons between analyses partially offset the extra DA cost. With current high-performance computing resources, hourly cycling is entirely feasible for regional domains of the size considered here.

\section{Conclusions}

We have assessed the impact of 3D-Var DA with RUC on short-range precipitation forecasting over Bali, Indonesia. Experiments with cycling intervals of 1, 3, 6, and 12~hours show a clear, systematic improvement in forecast skill as cycling frequency increases. The 1-hour configuration achieves temperature RMSE of 0.0--0.3$\,^\circ$C and precipitation RMSE of 1.295~mm~h$^{-1}$ at the Tabanan verification site---reductions of approximately 75\% and 57\%, respectively, relative to the Non-DA baseline.

Beyond these scalar metrics, hourly cycling also reproduces the spatial distribution of observed rainfall over western Bali far more accurately than coarser cycling, which tends to displace precipitation eastward. Independent radar verification confirms that high-frequency cycling captures convective reflectivity structures that lower-frequency cycling misses. Taken together, these results make a compelling case for RUC-based DA in nowcasting applications across the IMC.

Future extensions should include testing over additional events, seasons, and synoptic regimes. Assimilating radar reflectivity and radial velocity data has strong potential to further sharpen convective-scale forecasts. Hybrid ensemble--variational methods, meanwhile, may help overcome the limitations of static background error covariances in representing the flow-dependent errors that dominate at convective scales.

\section*{Acknowledgements}
This study was funded by ITB 3P Research Program 2026 (Top Tier Scheme) through the Directorate of Research and Innovation, Bandung Institute of Technology (Project ID: FITB.PN-6-219-2026). AWS and radar data were provided by the Indonesian Agency for Meteorology, Climatology, and Geophysics (BMKG). GFS data were obtained from the National Centers for Environmental Prediction (NCEP).

\section*{Author Contributions}
\textbf{NJT}: Conceptualization, Methodology, Supervision, Writing -- review \& editing.
\textbf{SHSH}: Investigation, Software, Formal analysis, Visualization, Writing -- original draft.
\textbf{IPFW}: Investigation, Software, Visualization, Writing -- original draft.
\textbf{FRF}: Software, Visualization, Writing -- review \& editing.
\textbf{RS}: Funding acquisition, Supervision, Resources, Writing -- review.
\textbf{DEI}: Funding acquisition, Supervision, Writing -- review \& editing.

\section*{AI Use Disclosure}
Claude 4.6 Sonnet (Anthropic, San Francisco, CA, USA) was used solely for English grammar and vocabulary editing. All scientific content---mathematical derivations, numerical simulations, data assimilation experiments, observational analysis, and interpretation of results---was performed entirely by the authors, who bear full responsibility for the work.

\end{document}